\documentclass[12pt]{iopart}

\usepackage{iopams}
\usepackage{graphicx}
\usepackage{bm}
\usepackage{setstack}
\begin{document}

\title{The three dimensionality of triple quantum dot stability diagrams}

\author{M.~C. Rogge and R.~J. Haug}
\address{Institut f\"ur Festk\"orperphysik, Leibniz Universit\"at
Hannover, Appelstr. 2, 30167 Hannover, Germany}

\ead{rogge@nano.uni-hannover.de}

\begin{abstract}
We present the full three dimensionality of an electrostatically
calculated stability diagram for triple quantum dots. The
stability diagram maps out the favored charge configuration of the
system as a function of potential shifts due to gate voltages. For
triple dots only a three dimensional visualization allows for the
complete identification of all its components. Those are most
notably the so called quadruple points where four electronic
configurations are degenerate, and quantum cellular automata (QCA)
processes. The exact positions of these features within the
stability diagram are now revealed. Furthermore the influence on
transport is studied by comparing the model with a two path triple
quantum dot made with local anodic oxidation. The two path setup
allows to study the influence of the dots arrangement.
\end{abstract}

\pacs{73.21.La, 73.23.Hk, 73.63.Kv}
\maketitle

\section{Introduction}

The electronic properties of quantum dots have been studied for
quite some decades by now \cite{Kouwenhoven-97}. Since quantum
dots have been proposed as crucial elements of future quantum
information technologies \cite{Loss-98}, these systems have gained
even more attention. First single quantum dots were studied as
they could be used to realize a quantum mechanical bit (qubit)
based on the electronic spin. Soon also double quantum dots
\cite{Wiel-03} came into focus. Due to technical and scientific
improvements during recent years, the research on quantum dots
finally reached a level allowing for the detailed analysis of
coupled triple quantum dots (e.g.
\cite{Vidan-04,Gaudreau-06,Schroer-07,Rogge-08a,Amaha-08}). They
represent the first step towards coupled qubit chains as they are
needed for quantum computers. They can also work as single qubits
themselves \cite{Hawrylak-05}. Other interesting applications can
be built with triple quantum dots, e.g. charge rectifiers and
ratchets \cite{Stopa-02,Vidan-04} or spin entanglers
\cite{Saraga-03}. They can be used to study other effects like
Kondo and Aharonov-Bohm effect \cite{Sakano-05,Kuzmenko-06} or
dark states \cite{Michaelis-06}.

The first and most crucial requirement for high level research on
quantum dot systems is the understanding of the stability diagram.
The stability diagram maps out the charge configuration of the
system as a function of potential detuning and does not depend on
the setup of transport leads (as long as there are no
instabilities, see below). In particular it shows degeneracy
points, where several charge configurations are degenerate, a
precondition for transport in multiple dot systems. For triple
quantum dots the full complexity of the stability diagram can only
be revealed in three dimensions. The degeneracy points consist of
at least four degenerate configurations (quadruple points).
Although any researcher is reliant on this map for exact analysis,
these diagrams are still not well known. Parts of it were
presented by Vidan et al. \cite{Vidan-05} and also Stopa
\cite{Stopa-02}, but the presentaions remain incomplete showing
only the first quadruple point. Gaudreau et al. \cite{Gaudreau-06}
firstly discovered several quadruple points with four degenerate
electronic configurations. Furthermore they found quantum cellular
automata (QCA) processes \cite{Lent-92,Amlani-99,Toth-01} leading
to charge rearrangements in two dots, when an electron is added to
the third dot. However, the diagram is presented in a two
dimensional visualization only, not capable to reveal the full
properties. Schr\"oer et al. \cite{Schroer-07} showed a three
dimensional visualization, but due to the large number of
electrons covered by the model, it is not possible to see the
dynamics at triple dot resonances, namely the positions and
properties of quadruple points. On vertical quantum dots Amaha et
al. \cite{Amaha-09} mapped out the stability diagram and also
presented it in three dimensions, but the quadruple points are not
resolved due to high temperatures. So even though the mathematics
is well established and stability diagrams were calculated,
measured and visualized, a complete and detailed three dimensional
stability diagram was never developed. This is a huge deficiency
in the understanding of these systems. As a result false
assumptions were made. For example Schr\"oer et al.
\cite{Schroer-07} stated that the number of quadruple points is
four. Furthermore, quadruple points do not feature four
intersecting charging or charge reconfiguration planes but six.
And most important, transport in a serial triple dot is possible
not only at these quadruple points.

\begin{figure}
 \includegraphics{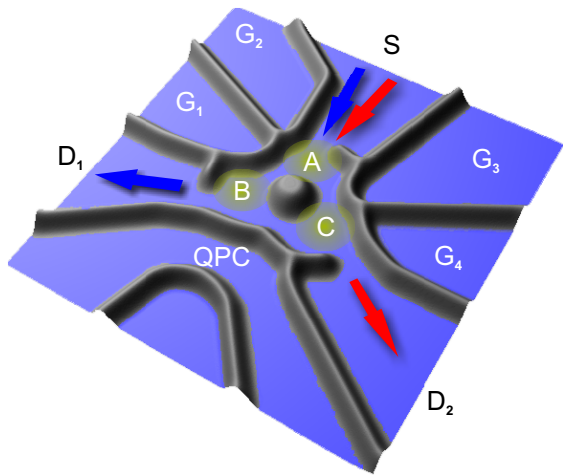}
 \caption{Artistic view of the measured triple quantum dot made with local anodic
 oxidation. Oxide lines (black) define three dots A, B, C coupled
 to Source (S), Drain1 (D1) and Drain2 (D2). The potentials are
 controlled via four gates G1, G2, G3 and G4. Transport is measured
 along two paths (blue and red arrows). A quantum point contact
 (QPC) serves for charge detection.}
 \label{fig1}
\end{figure}

In this paper we develop and visualize an electrostatic model for
stability diagrams of triple quantum dots. Although the underlying
mathematics is straight forward and does not differ considerably
from other ansatzes, the model and its visualization goes far
beyond existing models as it covers the full three dimensionality
of these diagrams. Thus it finally fills the former mentioned
deficiencies. It allows for the complete analysis of its
components especially the intensively discussed quadruple points
and the interplay with QCA processes. It is found, that there are
in general six quadruple points instead of only four. All six
points can show sequential transport. Furthermore, the QCA
processes appear not only at certain points or lines as implied
until now, but at a full plane in the center of the diagram.
Depending on the arrangement of the quantum dots and the leads,
the complete plane can contribute to transport.

The parameters for the model are chosen such that it compares to a
real triple quantum dot, measured with transport and charge
detection. This allows to study the influence of the model's
features on transport. The two path setup of the device allows to
understand the role of the dots arrangement. From the two path
transport a single path transport can be calculated that nicely
fits to the model's predictions.

\section{Electrostatic model}

For the calculation of the three dimensional model of a triple dot
charge diagram we determine the energetically favored electronic
configuration around the transition from zero electrons in the
system to one on each dot (equivalent to any other total electron
numbers from $N$ to $N+1$ on each dot) for any given set of three
gate voltages $V_{G1}$, $V_{G2}$ and $V_{G3}$ used for detuning
the potential. The number of possible electronic configurations of
a system containing $M$ quantum dots is $2^M$. The number of
configurations for a total electron number $N$ is ${M\choose N}$
with either zero or one electron on each dot. Thus a triple dot
has $2^3=8$ possible configurations $(N_A,N_B,N_C)$ with
$N_A,N_B,N_C$ being the electron numbers for the three dots named
$A$, $B$, $C$. There is one configuration for zero electrons in
the system: $(0,0,0)$, three for one electron: $(1,0,0)$,
$(0,1,0)$, $(0,0,1)$, three for two electrons: $(1,1,0)$,
$(1,0,1)$, $(0,1,1)$ and one for three electrons: $(1,1,1)$. The
electrostatic energies for these configurations as a function of
gate voltages are
\begin{flushleft}
for zero electrons:
\begin{eqnarray}
\nonumber E_0&=&0,
\end{eqnarray}
for one electron on dot $i$:
\begin{eqnarray}
\nonumber E_i&=&\frac{-e}{C_{\Sigma
i}}\left(C_{i1}V_{G1}+C_{i2}V_{G2}+C_{i3}V_{G3}\right),
\end{eqnarray}
for two electrons, one on dot $i$, one on dot $j$:
\begin{eqnarray}
\nonumber E_{i,j}&=&E_i+E_j+e^2\frac{C_{ij}}{C_{\Sigma i}C_{\Sigma
j}},
\end{eqnarray}
and for three electrons with one electron on dots $i,j,k$ each:
\begin{eqnarray}
\nonumber E_{i,j,k}&=&E_{i,j}+E_{i,k}+E_{j,k}-E_i-E_j-E_k,
\end{eqnarray}
\end{flushleft}
with $i,j,k\in \{A,B,C\}, i\neq j\neq k$. $C_{i1}$, $C_{i2}$,
$C_{i3}$ are the capacitances between dot $i$ and gates 1, 2, 3
and $C_{\Sigma i}$ is the total capacitance of dot $i$. The
capacitances $C_{ij}$ reflect the capacitive interaction between
dots $i$ and $j$, that leads to anticrossings in stability
diagrams. The energetically favored configuration of the system is
the one with the lowest energy depending on gate voltages. The
adjustment to the measured triple dot is done by adjusting the
capacitances.

\begin{figure}
 \includegraphics{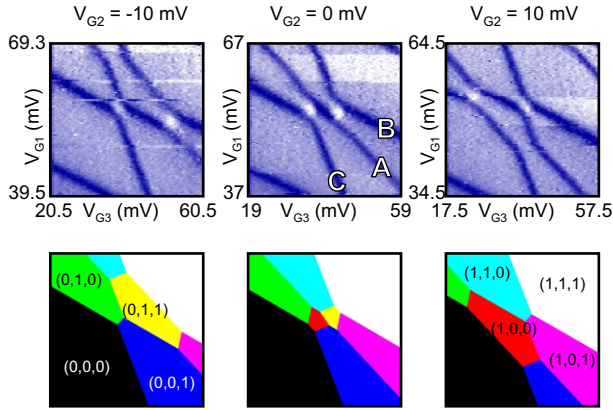}
 \caption{Upper part: charge detection as a function of $V_{G1}$ and $V_{G3}$ for three different voltages at
 G2. Dark lines correspond to charging of dots A, B, C (see middle
 plot). For $V_{G2}=-10$~mV and $V_{G2}=10$~mV anticrossings
 appear due to double dot physics, around $V_{G2}=0$~mV, triple dot
 resonances are formed. Lower part: electrostatic model for the
 upper cases using the following
capacitances: $C_{\Sigma A}=71.0$~aF, $C_{\Sigma B}=98.2$~aF,
$C_{\Sigma C}=35.6$~aF, $C_{A1}=6.2$~aF, $C_{A2}=5.15$~aF,
$C_{A3}=5.0$~aF, $C_{B1}=9.4$~aF, $C_{B2}=3.8$~aF,
$C_{B3}=4.05$~aF, $C_{C1}=1.5$~aF, $C_{C2}=1.1$~aF,
$C_{C3}=3.0$~aF, $C_{AB}=15.5$~aF, $C_{AC}=6.8$~aF,
$C_{BC}=4.4$~aF.}
 \label{fig2}
\end{figure}

An artistic view of the measured triple dot device is shown in
Fig. \ref{fig1}. The sample is made by local anodic oxidation
\cite{Ishii-95,Held-99,Keyser-00} on a GaAs/AlGaAs
heterostructure. The oxide lines (black) define three dots A, B
and C, which are connected to three leads Source (S), Drain1 (D1)
and Drain2 (D2). Thus transport can be measured along two paths,
from Source via dots A and B to Drain1 (blue arrows) and from
Source via A and C to Drain2 (red arrows). A quantum point contact
(QPC) is added for charge detection. The geometry of the device
and the measurement technique is explained in full detail in Refs.
\cite{Rogge-08a,Rogge-08b}.

The upper part of Fig. \ref{fig2} shows measurements on this
device recorded using charge detection. This method does not
depend on the setup of the leads (again as long as there are no
instabilities) and is thus capable to measure the stability
diagram. As a function of gate voltages $V_{G1}$ and $V_{G3}$
several lines are visible denoting charge rearrangements of the
system. Lines with a steep slope correspond to charging of dot C
with an additional electron, those with a shallow slope correspond
to charging of dot B. Intermediate slopes appear due to charging
of dot A (see central measurement). The three shown measurements
are cuts through the three dimensional stability diagram in the
vicinity of a resonance of all three dots. At $V_{G2}=-10$~mV this
resonance is not yet established. The system still acts as
independent double dots. Three different anticrossings are visible
that appear due to capacitive coupling for resonance of dots A-B,
A-C, and B-C, but those anticrossings are well separated. The same
description applies for $V_{G2}=10$~mV, except that the
anticrossings have swaped places. However, this is different at
the center image at $V_{G2}=0$~mV. Here all three anticrossings
have merged. In the vicinity resonances of a triple quantum dot
are expected, quadruple points form, QCA processes take place.

How the observed features can be understood in detail is examined
using the three dimensional model. The capacitances needed to fit
the model to the measurements are extracted from these and other
measurements. The result is shown in the bottom part of Fig.
\ref{fig2}. For the same voltages as in the measurements, the
energetically lowest electronic configurations $(N_A,N_B,N_C)$ are
plotted in a color encoded manner. With the chosen capacitances
(see caption of Fig. \ref{fig2}) the model fits pretty well.

\begin{figure}
 \includegraphics{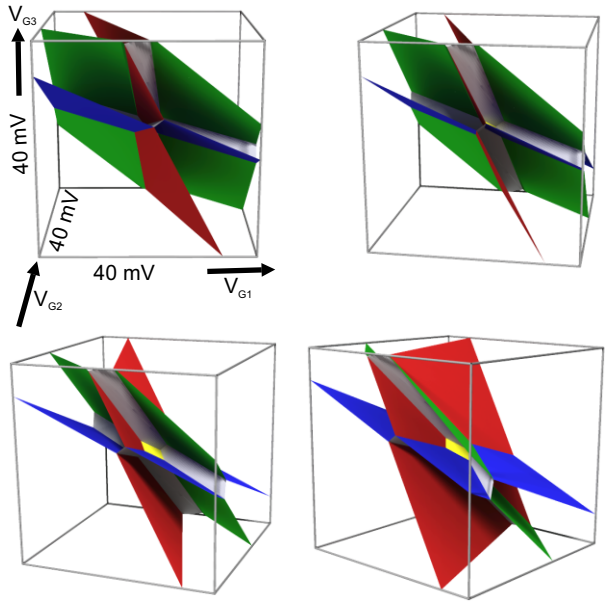}
 \caption{Three dimensional stability diagram shown from four perspectives.
 Three planes (green for charging of dot A, blue for B, red for C)
 cross in the center of the model and shift due to finite
 coupling. Quadruple points are formed and a central plane
 (yellow) featuring QCA processes.}
 \label{fig3}
\end{figure}

\section{Three dimensional Visualization}

The three images shown at the bottom of Fig. \ref{fig2} are as
explained cuts through the three dimensional stability diagram.
Even though the central image is a cut through the center of the
diagram, right here no quadruple points appear. However, they are
in direct vicinity, close enough to be visible in measurements due
to finite peak widths. To completely understand the model, a full
three dimensional visualization is indispensable. This
visualization is presented in Fig. \ref{fig3}. The stability
diagram is shown from four different perspectives as a function of
all three gate voltages. It is in general composed of three planes
colored red, blue and green. At each of these planes the total
charge of the system is changed by one electron which is added to
a certain dot via the leads. The green (blue, red) plane
corresponds to charging of quantum dot A (B, C). These planes
intersect as they have different gradients. At these intersections
double dot resonances are established. Due to the finite interdot
capacitances the dots interact and thus the planes shift.
Therefore when two planes intersect, anticrossings are formed with
three degenerate electronic configurations on each side. In two
dimensions these degeneracies are called triple points. In three
dimensions they are transformed into triple lines. A pair of
triple lines is connected with a grey plane. At these planes an
electron is moved from one dot to another without changing the
total number of electrons in the system. The spaces between all
the planes are characterized by stable electronic configurations.

\begin{figure}
 \includegraphics{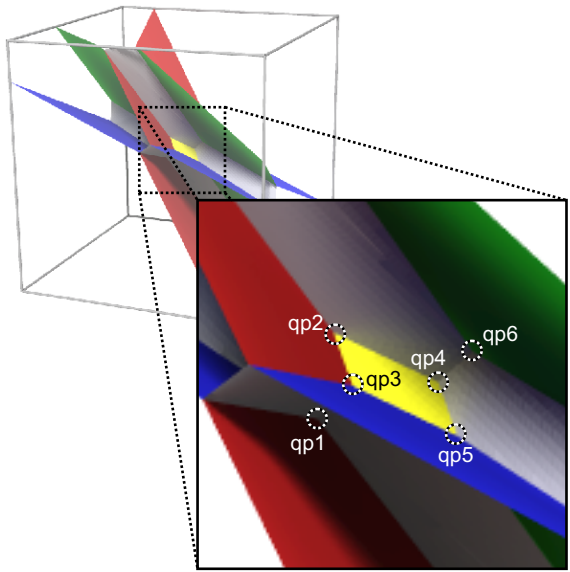}
 \caption{Highlighted center of the three dimensional stability diagram. The central yellow plane
 is defined by four quadruple points qp2, qp3, qp4 and qp5.
 Another one is below the plane (qp1) and above (qp6).}
 \label{fig4}
\end{figure}

At the center of the model the situation becomes more complex as
now all three colored planes intersect. Here quadruple points are
expected as well as QCA processes. Indeed the model shows a
feature that cannot be explained with double dot physics. In the
center of the model another plane appears (colored yellow). This
is shown in more detail in Fig. \ref{fig4} where the center of the
model is highlighted. The yellow plane is different to all other
planes of the model. It has a different gradient and does neither
correspond to charging of a certain dot only nor to pure charge
rearrangements within the system. Instead it features both. At the
whole plane the configurations (1,0,0) and (0,1,1) are degenerate.
Thus the following process is possible: when there is one electron
on dot A (1,0,0), a second electron can enter via dot B (C). At
the same time the electron on dot A is shifted to dot C (B)
leaving the system in the (0,1,1) configuration. This is a QCA
process, one of the major discoveries in triple dot systems. It
appears along the whole yellow plane.

This central plane is surrounded by points, where six planes
(colored and grey) intersect. These are the so called quadruple
points that feature four degenerate charge configurations.
Together with the QCA process, these points are the manifestation
of the triple dot. Six quadruple points exist, two more than
expected according to former publications. There is one below the
central plane (qp1), one above (qp6) and four more that actually
define the central plane (qp2, qp3, qp4, qp5). Only under special
conditions, the number of degeneracy points can be reduced. If for
example the three dots do not interact, there is only one point,
where all planes intersect with all eight charge configurations
being degenerate. For the more general case of six quadruple
points, four of them are already known. For those the following
configurations are degenerate:
\begin{flushleft}
qp1: (0,0,0), (1,0,0), (0,1,0), (0,0,1),\\
qp2: (1,0,0), (0,1,0), (1,1,0), (0,1,1),\\
qp5: (1,0,0), (0,0,1), (1,0,1), (0,1,1),\\
qp6: (1,1,0), (1,0,1), (0,1,1), (1,1,1).
\end{flushleft}
The two quadruple points qp3 and qp4 with degenerate
configurations
\begin{flushleft}
qp3: (1,0,0), (0,1,0), (0,0,1), (0,1,1),\\
qp4: (1,0,0), (1,1,0), (1,0,1), (0,1,1),
\end{flushleft}
were not known so far, as they appear in a different way in two
dimensional cuts through the stability diagram. For the other
quadruple points even a two dimensional cut shows all four charge
configurations that meet at the quadruple point. For qp3 and qp4,
only three configurations meet at one point in two dimensions. The
fourth configuration touches this point coming from the third
dimension. So this point does not appear as a quadruple point in
two dimensions. qp3 and qp4 were therefore overlooked.

\begin{figure}
 \includegraphics{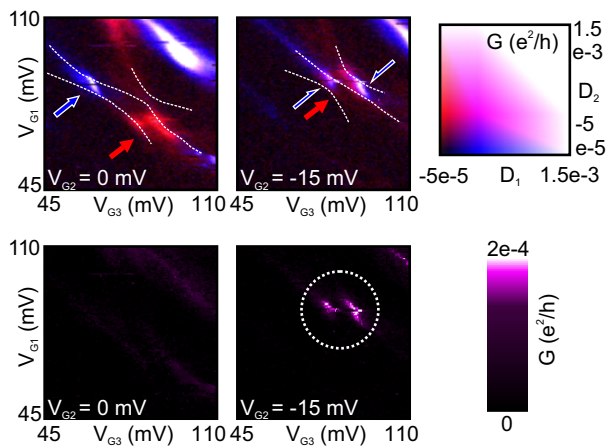}
 \caption{Upper part: Conductances along the two paths (path1 in blue, path2 in red) as a function of $V_{G1}$ and
 $V_{G3}$ off (left) and on (right) triple dot resonance. Features
 from double and triple dot physics are visible. Lower part: From
 measured transport calculated hypothetic serial conductance. Double
 dot features vanish, only triple dot resonances remain. They appear as a two spot feature (dashed circle).}
 \label{fig5}
\end{figure}

\section{Comparison with transport measurements}

As mentioned before, the properties of the stability diagram (and
of charge detection) do not depend on how the dots are connected
to the leads. In particular they do not depend on the transport
paths. This is not true for the conductance properties. They
strongly depend on the dots' arrangement. With the model now well
known the transport properties of our two path triple quantum dot
can be analyzed. The upper part of Fig. \ref{fig5} shows two
transport measurements as a function of $V_{G1}$ and $V_{G3}$. The
conductance of both transport paths is plotted simultaneously,
path1 in blue, path2 in red (results from charge detection are
added as dashed lines). The left plot at $V_{G2}=0$~mV corresponds
to the back side of the three dimensional model. There is no
resonance of all three dots. Nevertheless features are visible in
both paths showing double dot resonances. The red features (marked
with a red arrow) appear due to a resonance of dots A and C. They
correspond to the triple lines on both sides of the anticrossing
when the model's red and green planes intersect (see Fig.
\ref{fig2}). Indeed one can see that the two involved triple lines
form a double feature of two triple points in the measurement. The
blue feature (blue arrow) corresponds to the intersection of the
blue and the green planes from the model due to resonance of dots
A and B. Although there are two triple lines involved as well,
there is only one spot visible as the anticrossing is not large
enough and cannot be resolved in transport. The right plot at
$V_{G2}=-15$~mV corresponds to the center of the model. Here
resonances of all three dots appear. This becomes obvious as both
paths show transport at the same time. Quadruple points must be
involved. But, the exact spots of triple dot resonances are not
well visible, as the measurement still shows features from double
dots. However, due to the two path setup, there is an elegant way
to seperate the triple dot features.

Assuming the two paths to be ohmic with conductances $G1$ and
$G2$, one can calculate a hypothetic transport from Drain1 via
dots B, A, C to Drain2. Thus the arrangement of the dots can be
mathematically changed to form three dots in series with a
combined conductance $G=\frac{1}{1/G1+1/G2}$. Such arrangement can
only show transport, if all three dots are resonant. Therefore,
double dot features should be wiped out. Of course this method
does not include nonohmic effects. Furthermore the tunnel
resistance to the Source is still included which is not the case
in a real serial device. And finally the real arrangement is of
course not changed. The observed features must still be analyzed
in terms of the two path setup.

The result of this method is presented in the lower part of Fig.
\ref{fig5} for both voltages $V_{G2}$. Indeed the double dot
features are strongly suppressed. For $V_{G2}=0$~mV almost no
conductance is visible. All the triple points have disappeared.
For $V_{G2}=-15$~mV transport is suppressed as well. However, two
spots are left (dashed circle). These are the features from the
triple dot, that appear now well separated.

As the conductance properties depend on the dots' arrangement, it
is not guaranteed that all quadruple points can contribute to
transport. However, the quadruple points qp1 and qp6 are always
visible as long as there are at least two leads connected to three
coupled dots. At qp1 an electron can enter the empty system at any
dot that is connected to a lead. It can then travel from dot to
dot until it reaches the exit lead. The same accounts at qp6 for
holes. Quadruple points qp2 to qp5 are not that trivial. For the
two path setup, transport means charge exchange between the Source
lead and one or both Drain leads. This is possible at qp2 and qp5.
At qp2, transport in path1 can occur via the transition (1,0,0)
$\rightarrow$ (1,1,0) $\rightarrow$ (0,1,0) $\rightarrow$ (1,0,0).
An electron can enter from Drain1 to dot B, another electron then
leaves via dot A to Source and the remaining electron shifts from
B to A. The analoge process is possible for a hole in path2 with
the transition (0,1,1) $\rightarrow$ (0,1,0) $\rightarrow$ (1,1,0)
$\rightarrow$ (0,1,1). Similarly transport is possible at qp5 in
both paths via (0,1,1) $\rightarrow$ (0,0,1) $\rightarrow$ (1,0,1)
$\rightarrow$ (0,1,1) in path1 and via (1,0,0) $\rightarrow$
(1,0,1) $\rightarrow$ (0,0,1) $\rightarrow$ (1,0,0) in path2. At
qp3 and qp4 no transport is possible for the two path setup. At
both quadruple points neither electrons nor holes can enter the
system via Source. So the observed features in Fig. \ref{fig5}
must be quadruple points qp1, qp2, qp5 and qp6. According to the
model, the quadruple points qp1 and qp2 are much closer than e.g.
qp1 and qp5. The same accounts for qp5 and qp6. So there are two
pairs of quadruple points. Indeed the measurements shows only two
spots instead of four. They correspond to these two pairs. The
quadruple points within a pair are again too close to be resolved.
The left spot contains qp1 and qp2, the right spot contains qp5
and qp6. The observation of these pairs confirms the model nicely.

The gap between the two pairs also confirms that the real dot
arrangement is of course not changed by the calculation of a
hypothetic serial transport. For a real serial system with Source
and Drain at dots B and C, transport would occur along the whole
central plane via the QCA process (and via sequential tunneling at
qp3 and qp4). As this process does not involve charge exchange
with the Source lead, it does not produce transport in our device.

Even these stability diagrams can have instabilities. If for
example there is only one lead attached to dot A at one end of a
triple dot chain A, B, C, the configuration (1,0,0) could not be
transferred directly to (0,0,1) even if both were degenerate. One
had to lift this degeneracy to transfer the electron from A to B
and from B to C. Then the degeneracy could be established again
with the system in (0,0,1). So the actual charge configuration
depends on the path through the stability diagram. Such
instabilities could appear in quadruple quantum dots even when two
leads are attached.

\section{Conclusion}

In conclusion we have visualized the full three dimensionality of
triple dot stability diagrams. We have developed an electrostatic
model that was fit to match the properties of a triple dot device
with a two path setup. Due to the full visualization we were able
to completely analyze all the model's features for triple dots. We
found the maximum number of quadruple points (and degeneracy
points in general) to be six instead of four as reported in other
publications. We identified quantum cellular automata processes
that occur along a whole plane in the center of the model. The
comparison with the two path triple dot confirms the model and is
capable to explain all features visible in transport. Due to the
two path setup we were able to separate double dot features from
triple dot features. As a result, the transport properties in
triple dots strongly depend on the dot arrangement. Under certain
circumstances transport is possible not only at quadruple points
but along the whole central plane using the quantum cellular
automata process.

\section*{Acknowledgements}

For the heterostructure we thank M. Bichler, G. Abstreiter, and W.
Wegscheider. This work has been supported by BMBF via nanoQUIT.




\begin{thebibliography}{10}



\bibitem{Kouwenhoven-97} Kouwenhoven L~P et al. in {\it Mesoscopic Electron Transport}, edited by Sohn L~L, Kouwenhoven L~P and Sch\"o{}n G, Kluwer, Dordrecht, 1997 vol. 345 of Series E, pp. 105--214.

\bibitem{Loss-98} Loss D and DiVincenzo D~P 1998 {\it Phys. Rev. A} \textbf{57} 120.

\bibitem{Wiel-03} van~der Wiel W~G et al. 2003
{\it Rev. Mod. Phys.} \textbf{75} 1.

\bibitem{Vidan-04} Vidan A, Westervelt R~M, Stopa M, Hanson M and Gossard
A~C 2004 {\it Appl. Phys. Lett.} \textbf{85} 3602.

\bibitem{Gaudreau-06} Gaudreau L et al. 2006 {\it Phys. Rev. Lett.} \textbf{97} 036807.

\bibitem{Schroer-07} Schr\"oer D et al. 2007
{\it Phys. Rev. B} \textbf{76} 075306.

\bibitem{Rogge-08a} Rogge M~C and Haug R~J 2008
{\it Phys. Rev. B} \textbf{77} 193306.

\bibitem{Rogge-08b} Rogge M~C and Haug R~J 2008
{\it Phys. Rev. B} \textbf{78} 153310.

  \bibitem{Amaha-08} Amaha D et al. 2008
{\it Physica E} \textbf{40} 1322.

\bibitem{Hawrylak-05} Hawrylak P and Korkusinski M 2005
  {\it Solid State Commun.} \textbf{136} 508.

\bibitem{Stopa-02} Stopa M 2002
{\it Phys. Rev. Lett.} \textbf{88} 146802.

\bibitem{Saraga-03} Saraga D~S and Loss D 2003
{\it Phys. Rev. Lett.} \textbf{90} 166803.

\bibitem{Sakano-05} Sakano R and Kawakami N 2005{\it Phys. Rev. B} \textbf{72} 085303.

\bibitem{Kuzmenko-06} Kuzmenko T, Kikoin K and Avishai Y 2006
{\it Phys. Rev. Lett.} \textbf{96} 046601.

\bibitem{Michaelis-06} Michaelis B, Emary C and Beenakker C~W~J 2006
{\it Europhys. Lett.} \textbf{73} 677.

\bibitem{Vidan-05} Vidan A, Westervelt R~M, Stopa M, Hanson M and Gossard
A~C 2005 {\it J. Supercond.} \textbf{18} 223.

\bibitem{Lent-92} Lent C~S, Tougaw P~D, Porod W and Bernstein G~H 1992
{\it Nanotechnology} \textbf{4} 49.

\bibitem{Amlani-99} Amlani I et al. 1999
{\it Science} \textbf{284} 289.

\bibitem {Toth-01} Toth G and Lent C~S 2001
{\it Phys. Rev. A} \textbf{63} 052315.

\bibitem{Amaha-09} Amaha D et al. 2009
{\it Appl. Phys. Lett.} \textbf{94} 092103.

\bibitem{Ishii-95} Ishii M and Matsumoto K 1995
{\it Jpn. J. Appl. Phys.} \textbf{34} 1329.

\bibitem{Held-99} Held R, L\"uscher S, Heinzel T, Ensslin K and Wegscheider W 1999
{\it Appl. Phys. Lett.} \textbf{75} 1134.

\bibitem{Keyser-00} Keyser U~F, Schumacher H~W, Zeitler U, Haug R~J and Eberl K 2000
{\it Appl. Phys. Lett.} \textbf{76} 457.

\end{thebibliography}

\section*{References}

\end{document}